\documentstyle[aasnorcv,11pt,references,hacks]{article}
\begin{document}
\baselineskip 1.2pc	 
%
%
\input BoxedEPS.tex
\input BoxedEPS.cfg
\HideDisplacementBoxes

\bgroup\parindent=0pt\parskip=\itemsep
\def\refpar{\par\hangindent=3em\hangafter=1}}
\def\endselectedbib{\refpar\egroup}
\def\pageassign#1#2{[pages: #1, author: #2]}


\title{Cosmic Microwave Background Observations in the Post-Planck Era}

\author{
J.~B. PETERSON, 
J.~E. CARLSTROM, 
E. S. CHENG,
M. KAMIONKOWSKI, 
A.~E. LANGE, 
M. SEIFFERT,
D.~N. SPERGEL, 
A. STEBBINS
}


Electronic mail: jbp@cmu.edu


\begin{abstract}

{\baselineskip 1.2pc	

The Microwave Anisotropy Probe and Planck missions will
provide low noise maps of the
temperature of the cosmic microwave background (CMB). 
These maps will allow measurement of the power spectrum of
the CMB 
with measurement noise below cosmic variance for
$\ell < 1500$. It is anticipated that no further all sky CMB  
temperature observations will be needed after Planck.
There are, however, other CMB measurements for which
Planck will be not the end but the beginning.
Following Planck, precise CMB {\it polarization} observations will offer the
potential to study physical processes at energies as high as $10^{19}$ GeV.
In addition, {\it arcminute scale}, multi-frequency observations will allow 
study of the early phases of the formation of large-scale structure in the universe.
}

\end{abstract}


\section{ Scope of Report }

The Cosmic Microwave Background Future Missions Working Group (the authors of this report)
was created 
by NASA headquarters to consider new missions to follow
a successful 
Planck mission. The discussion was  restricted to missions of cost greater than
Explorer Class missions (\$200M).  This report presents the conclusions of the
working group, arrived at after extensive discussions involving
the CMB and high energy physics communities.
This report is not intended to be a complete review of the literature on CMB
science; a few references have been included, however, to serve as starting
points for more thorough research.

\section{ Introduction }

The development of the hot-big-bang model for the early history of the universe 
is one of the crowning achievements of twentieth century science.    
It is remarkable that we may now understand what was going on in the universe 
just one second into the big bang expansion.
A cornerstone in our understanding of modern cosmology has been the
information gathered from measurements of the Cosmic Microwave Background (CMB) 
radiation. Just now, at the close of the twentieth century, an important piece of the
big-bang puzzle is about to be put into place. Preliminary CMB
results announced recently
indicate that of the three possible geometries of the universe---open, flat,
or closed---ours appears to be flat ($\Omega_{total}=1$). 
At the same time, measurements of the flux from distant supernovae indicate that 
the universal expansion may actually be accelerating, that Einstein's 
cosmological constant $\Lambda$ may be significant.
These are exciting times: measurements of the cosmological parameters
are about to tell us the ultimate fate of the universe; meanwhile observations
are confronting us with mysteries such as that of an accelerating universe.
Soon, two
CMB satellite missions already in
the NASA and ESA pipelines (MAP and Planck) are expected to
provide precise measurements of $\Omega$ and $\Lambda$. 
In addition,
these experiments 
will rigorously test whether the large structures 
we find in the universe today, such as
galaxy clusters and superclusters, 
grew from a nearly scale-invariant spectrum of tiny primordial
density  perturbations. They will test the idea that the largest 
structures in the universe began as minute quantum mechanical
fluctuations during an inflationary epoch in the very early Universe.
MAP and Planck are extraordinarily capable missions, but are they {\it ultimate} 
CMB missions? Are there important issues
these two spacecraft do not address? That is the subject of this report. (For more 
on this issue, see also \cite{halp99}.)

CMB images are maps of ancient temperature structure.
A precise measurement of the CMB energy spectrum, carried out using the FIRAS instrument on COBE
(\cite{Fixetal94}), 
showed that the energy spectrum closely follows a Planck curve.  Because
of this,
the CMB is today widely accepted to be the thermal
radiation relic of the hot big bang explosion.  Under this interpretation  we know that
most cosmic background photons have traveled freely to us from a last scattering 
that occurred when the universe first became de-ionized, just 300,000 years after the 
start of the big bang explosion. The
mapping of CMB sky structure therefore amounts to the measurement of small temperature
variations that were present in the universe at that early time.

CMB observations test gravitational collapse models of structure formation.
Using DMR on COBE (\cite{Smoetal92}; \cite{kogut95}) the level of CMB sky structure has been measured
from
10 to 180 degrees.
The sky structure detected
falls in line with measurements 
of large-scale structure in the distribution of galaxies in the 
universe today.  It is therefore generally 
accepted that at ten degree scales the small temperature
differences on the sky measured using  CMB telescopes are the result of  weak 
gravitational potential variations (and therefore density variations)
in the early universe (\cite{SacWol67}; \cite{bennett96}; for a
review, see \cite{WhiScoSil94}).  These weak potential wells
acted as seeds for the growth, by
gravitational collapse, of the network of galaxy sheets and voids that we find in 
the universe today. Thus, CMB observations also provide a crucial test of
the structure-growth-by-gravitational-collapse paradigm, a key tenet of cosmological
theory.

\ForceWidth{0.75\textwidth}
\begin{figure}
\begin{center}
\BoxedEPSF{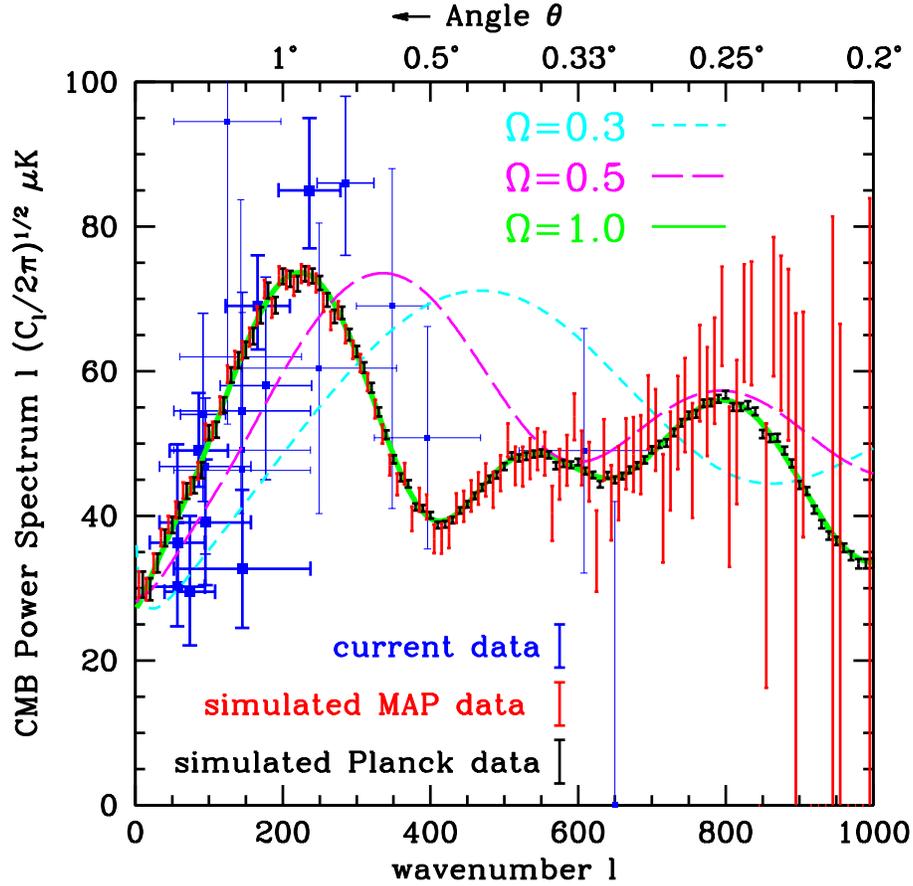}
\end{center}
\caption[Spectrum Plots]{{\bf Spatial Power Spectrum of CMB Sky Structure.\ \ } 
Shown in this figure (taken from \cite{kamionsci98})
are: current CMB power spectrum measurements, plotted as squares with
error bars; three theoretical predictions of the power spectrum, plotted
as curves; and projected error limits for MAP and Planck, plotted as
red and black vertical error bars.
The use of a spacecraft will
allow measurements of CMB sky structure with precision well beyond that 
possible from the ground.  Because of that precision observations with 
MAP and Planck are expected to determine
the values of the cosmological parameters.
\label{fig:spectrum} }
\end{figure}

MAP and Planck should tell us the fate of the universe.
As seen in Figure \ref {fig:spectrum}, at degree angular scales 
there is now good evidence that CMB sky structure is
present at an amplitude about three times as large as that detected with
COBE at large angular scales.
That amplification is expected; it  occurs  naturally as a result of  acoustic oscillations
that took place while the universe was ionized, before age 300,000 years.
These oscillations and the sky structure they produce are sensitive
to the matter density in the universe today and to the other parameters of the
cosmological world model.  Measurement of degree angular scale
CMB sky structure, at the sensitivity offered using the MAP and Planck spacecraft,
is expected to provide, with percent-level precision, a measurement of
the density of the universe $\Omega_{total}$, 
the expansion rate  H$_\circ$, 
the cosmological constant $\Lambda$, 
as well as a measurement of the fractions of 
material  present today
in the forms of hot and cold dark matter.  
By accurately determining
these cosmological parameters, CMB observations made with MAP and Planck 
have the 
potential to
tell us the ultimate fate of the universe.  Of course,
MAP or Planck may find sky structure that is completely different from
that expected.  If this occurs, we may not be able to extract the cosmological parameters
from the CMB data; instead, MAP and Planck will have presented for discussion an exciting
new puzzle.

\smallskip
\smallskip
\centerline{\bf Beyond MAP and Planck}
Following Planck there are two interesting new directions for CMB observations.
A sensitive CMB polarization experiment can be used to detect
the primordial gravitational waves produced during inflation.
In addition, fine scale observations 
can provide images of the largest structures in the universe
just as they were beginning to dissipate the heat of their gravitational collapse.

Inhomogeneities in the universe that were present at age 300,000
years were themselves the result of earlier processes.  By studying degree
angular scale CMB structure we learn about processes, inflation
for example, that may have produced gravitational potential perturbations.
The careful study of CMB structure, in particular degree angular scale 
{\it polarized} sky structure, can be used to detect not just potential perturbations
but also gravitational waves (\cite{BonEfs84};
\cite{polnarev85}; \cite{KamKosSte97a}; \cite{Zal97}; \cite{keating98})
Unlike photons, these gravitational waves travel freely through
the ionized early universe, so the study of CMB polarization
may allow us to look much further back in time--and at much higher
particle energies--than has been possible before.
However, the sensitivity needed for these observations
exceeds that attainable with the Planck satellite.
As described in the Polarization section below,
if a new instrument can be built with sufficient sensitivity and
with sufficient control of systematic errors, the
resulting observations have the potential to allow 
the study of physics at energy scales as high as $10^{19}$ GeV,  
far beyond the reach of the largest feasible
particle accelerators (see, e.g., \cite{KamKos99} for a recent review).

On angular scales of one arcminute and finer, the CMB picks up an impression from 
the material it passes through. Since the CMB passed through the dark
ages of cosmology, before stars or quasars formed, the study of arcminute 
scale CMB features
is the only technique we know for 
viewing the very early processes of structure formation. Using the CMB as
an intense, very high redshift backlight, we can study the early history of 
galaxy clusters and we can witness their gradual acceleration as they began
to fall into the deepening potential wells around them. We should also 
be able to view directly, at the time of first formation, the precursors
to the 100 Mpc scale sheets (e.g. the ``Great Wall'') we find in the galaxy 
distribution today.
No other known cosmological
observable can provide such direct information on the great 
structures as they were forming.  Because of the 
high angular resolution required, this work exceeds the capabilities of
the MAP and Planck missions. These observations are described in the Fine Scale section below.


\section{ Polarization }

\begin{figure}
\begin{center}
\BoxedEPSF{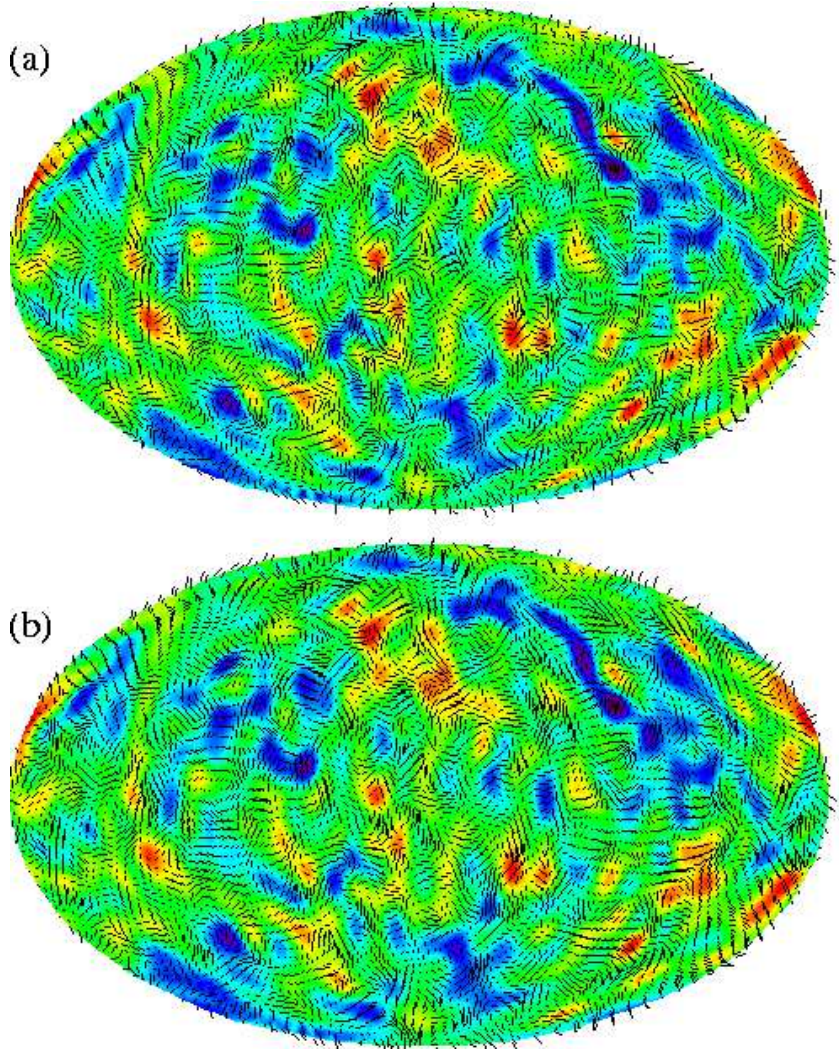}
\end{center}
\caption[CMB Maps With Polarization Vectors]{{\bf CMB Map With Polarization Vectors\ \ }
Shown above (from \cite{caldwell98}) are two simulated maps of CMB temperature structure (colors) and 
polarization fraction (bars).  The temperature structure in both maps is 
the same.  The polarization vectors in panel (a) were calculated assuming 
polarization was produced by gravitational waves during inflation at a  high 
energy.  The polarization vectors in panel (b) were calculated assuming
polarization was produced by density perturbation due to inflation at a lower
energy.  The polarization vector field in (a) has curl, but the field in (b) 
does not.  Therefore a precise measurement of the curl of the polarization
vector field of the CMB will expose or disprove the existence of gravitational 
waves, and constrain the energy scale of inflation.  
\label{fig:both} }
\end{figure}

Polarization of the CMB contains a wealth of 
information on primordial perturbations that will
not be provided by the temperature map.  
It was recognized early on that CMB sky structure should be polarized
(\cite{rees68}, \cite{kaiser83}), and theoretical studies of polarization
continue.  For a
tutorial on CMB polarization see \cite{HuWhi97b}. Using temperature information alone it will not 
be possible to confidently separate the three classes of initial
perturbations: density, pressure, and gravitational waves.  An example of this
degeneracy is shown in Figure \ref{fig:both}. Two simulated sky maps of temperature
and polarization are shown.
In one case the sky structure is due to gravitational waves,
in the other case the structure is due to density perturbations.
These two temperature maps differ by less than
the cosmic variance so no CMB temperature observation can ever
distinguish between them.  
Pressure perturbations can also create temperature structure indistinguishable
from the two maps shown in Figure \ref{fig:both}.  
Fortunately, 
the three sources of primordial perturbations produce
distinctly different polarization patterns.  
Polarization information can break the perturbation-type degeneracy.

\ForceWidth{.7\textwidth}

\begin{figure}
\begin{center}
\BoxedEPSF{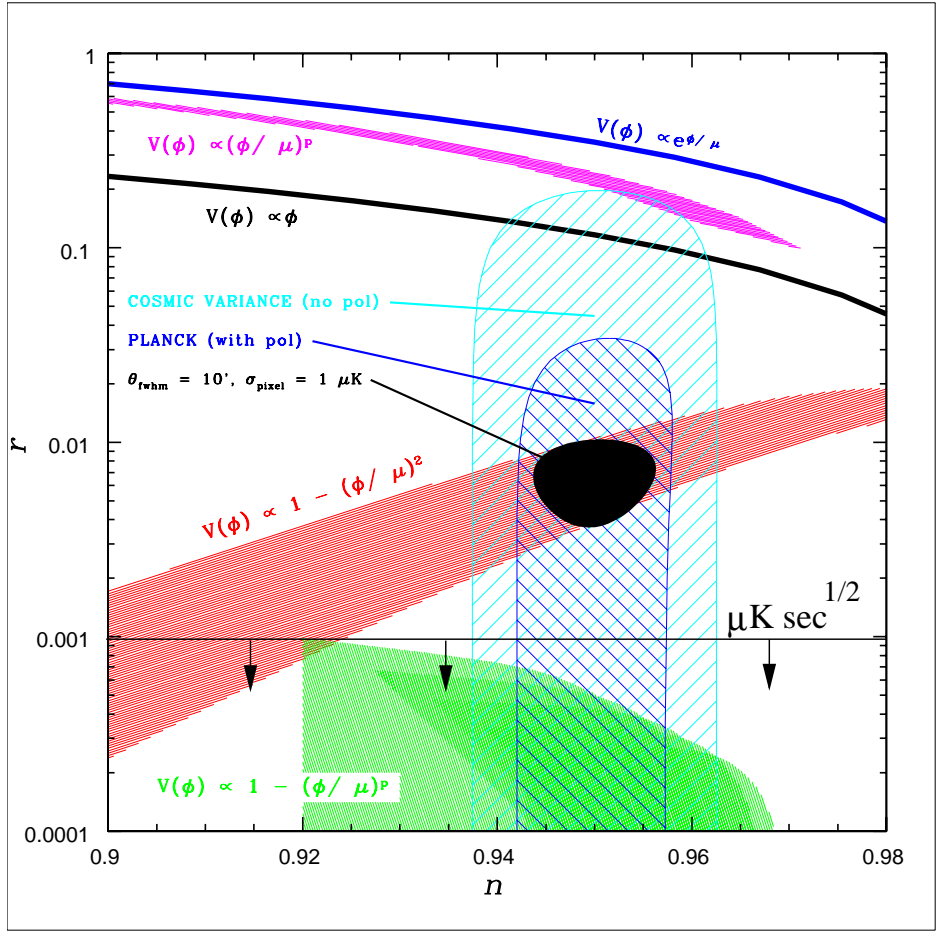}
\end{center}
\caption[Tests of inflation models]{{\bf Tests of inflation models.\ \ }
Five types of inflation models are shown in {\it r} (tensor/scalar ratio) vs
{\it n} (spectral tilt index) space. In this simulation it is assumed
that the true values of {\it r} and {\it n} fall in the center of
the black region.  The outer hatched region shows the
constraint in this space possible with noiseless temperature maps, but without polarization
information.  The inner hatched region is the constraint possible with the
currently planned Planck capabilities.  The black region shows the constraint possible
with a polarization instrument with sensitivity three times better than Planck. (From 
\cite{kinney98}.)
\label{fig:kinney} }
\end{figure}

Polarization information also allows fine selection among inflation models.
Figure \ref{fig:kinney}, from \cite{kinney98}, shows the selective power of polarization 
information.  Five different inflation models are shown.  Without polarization information
it is not possible to distinguish the models.  However, a polarization 
experiment with a sensitivity three times better than Planck
would allow a strong selection.

\ForceWidth{0.75\textwidth}

\pagebreak
\centerline{\bf Gravitational Waves from Inflation}
It is particularly exciting that 
gravitational waves are potentially detectable through measurement of CMB polarization.
The clouds of ionized gas viewed with CMB telescopes were in motion
at the time of last scattering.  These motions produce patterns in the
polarization in the CMB. Fractional polarization of CMB 
sky structure is expected at the few percent level in all models 
of the universe. 
In models with strong gravitational waves, however, swirling
motions in the ionized gas produce polarization patterns that
differ from gravitational
infall patterns due to density perturbations.

A map of polarization vectors on the sky can be
decomposed into a curl and a curl-free part.  Density
perturbations produce scalar perturbations to the spacetime
metric.  Since density perturbations have no handedness, they cannot produce a
curl.  On the other hand, gravitational waves are tensor
perturbations.  They can have a handedness (there are right- and
left-handed circularly-polarized gravitational waves), so they
do produce a curl.  Thus, a curl component in the polarization
provides a smoking-gun signature for a gravitational-wave
background (\cite{KamKosSte97a}; \cite{Zal97}).

We will only understand the big bang when we also understand
high energy physics.
The  current hot big bang
model
accounts for the universal expansion, the presence
and energy spectrum of the CMB, the light-element
abundances, and can certainly accommodate primordial
perturbations, but the hot big bang model leaves many questions unanswered.  For
example, why is the universe flat?  Why is it so smooth? Why was 
there a big bang in the first place? And where did the
primordial inhomogeneities mapped with CMB telescopes come from?  
Just as the light element abundances in today's universe could only be 
understood through an understanding of the underlying nuclear physics, 
we now appreciate that the answers to these
current questions must be accompanied by 
development of our understanding of particle theory at very high energy scales.
The answers to these questions will probably not be obtained without
a concomitant understanding of the unification of the fundamental
forces of Nature.

Inflation, a proposed period of accelerated expansion in the early
universe driven by the release of vacuum energy, offers
answers to all the questions listed above. If MAP
and Planck confirm that the universe is flat and if the spatial
power spectrum of the CMB agrees with that expected from 
adiabatic initial perturbations,
we may indeed  be on the right track with inflation. If so,
we must further test the inflationary
hypothesis and attempt to determine the physics responsible for
inflation.

By precisely mapping CMB polarization we may be able to determine the energy scale of
inflation.
There is a very broad range of models of inflation that
produce a flat universe and a nearly scale-invariant
spectrum of primordial density perturbations. 
A priori, inflation could have occurred when the 
universe had a temperature anywhere from roughly electroweak scale, $100\ {\rm
GeV/k}$ to the Planck temperature, $10^{19}\,{\rm GeV/k}$. 
However, it may be possible to determine the cosmic temperature at the end
of the era of inflation because
inflation also generically predicts the existence of 
a stochastic background of
long-wavelength gravitational waves (\cite{AbbWis84}). The amplitude of this 
gravitational-wave background
is proportional to the square of the inflationary energy scale. 
Detection of these gravitational waves would therefore
provide a direct measurement of the energy scale of inflation
and point to the new physics responsible
for inflation. 
If inflation occurred at very high energies the resultant gravitational waves 
would have produced large amplitude
temperature structure in the CMB. 
But the COBE sky structure is weak ($\Delta T/T \sim 10^{-5}$),
so  COBE data can already be
used to constrain the energy scale of inflation to be less 
than $\sim 10^{16}~{\rm GeV}$. However, since CMB temperature
structure from gravitational waves cannot
be unambiguously distinguished from that due to density
perturbations, we don't know right now which produced the
CMB structure that we have detected. Polarization maps will be needed to 
settle this issue.
 
If inflation has something to do with GUTs (unified theories of
the electroweak and strong interactions), as most theorists
surmise, then 
the inflation era probably occurred at temperature $\sim 10^{16}$ Gev and 
the gravitational-wave signal is 
detectable with a next-generation CMB polarization
experiment.  Detection of such a signal would be truly
extraordinary: It would (1) allow us to penetrate the
last scattering surface, giving us a glimpse of the universe
as it was at a time $10^{-38}$  seconds after the initial
singularity; (2) provide a window to new physics at energy
scales more than 12 orders of magnitude beyond those accessible at
accelerator laboratories; (3) explain the origin of primordial
perturbations. And, since gravitational waves and density
perturbations are produced during inflation by an
analog of Hawking radiation, detection of a gravitational-wave
background would (4) be the first observation of the
effects of quantum field theory in curved spacetimes, and thus
would provide clues to the nature of quantum gravity.

An observable polarization-curl signal is not guaranteed even if
inflation did occur (it may have taken place at a lower energy
scale), but the signal should be detectable if inflation took
place near the GUT scale. 
This means that even a null result from a sensitive polarization experiment would
be quite interesting. It would indicate that inflation did {\it not}
arise from a GUT phase transition or from quantum gravity effects which also
place the inflation epoch at a high temperature.

High-sensitivity polarization 
maps will be used to study a wide range of interesting cosmological
physics in addition to probing gravitational waves.  For
example, the polarization pattern can be used to
isolate the peculiar-velocity contribution to degree-scale
temperature anisotropy. This information will
be essential to  discriminate between
various structure-formation models that give rise to the same
temperature perturbations.  
Polarization information can also be used to constrain the ionization history of
the universe and thus probe the earliest epochs of formation of
gravitationally-collapsed objects in the universe.  Polarization 
can also probe primordial magnetic fields. Perhaps most importantly,
polarization data may contain surprises due to unanticipated
physical processes.

COBE provided valuable
information on the origin of large-scale structure.  It
confirmed the notion that large-scale structure grew via
gravitational infall from primordial density perturbations. Also,
the COBE data now provide a  normalization for the power spectrum of
primordial perturbations in any gravitational growth model.  MAP
and Planck are designed to accurately measure the 
CMB temperature
structure expected on the sky and thereby provide precise information on the
primordial spectrum of density perturbations.
Although both MAP and Planck will provide some polarization information
they lack the sensitivity to detect the gravitational wave signal from
GUT scale inflation.  
To complete our study of the primordial fluctuations
we must also map the CMB polarization to small
angular scales with a sensitivity
at
the cosmic-variance limit in the same way that Planck will map
the temperature structure.

\smallskip
\smallskip
\centerline{\bf Specifications of a Polarization Experiment}
Angular resolution:
Although gravitational-wave-produced
polarization structure in the CMB occurs over a range of 
angular scale from 100 to 0.1 degrees, angular resolution finer than one degree is required
to see the predicted turnover in the power spectrum (\cite{kamionkowski98}). Detecting this
turnover will
be valuable in discriminating between gravitational waves and an
unsubtracted foreground (whose power spectrum would rise at smaller angular 
scales). A polarization experiment should
have an angular resolution of 0.3$^\circ$.

Sensitivity:  As mentioned above, the amplitude of the
gravitational-wave background---and therefore, the amplitude of
the gravitational-wave-induced curl component of the
polarization---depends on the energy scale of inflation, which
is currently unknown.  If inflation has something to do with
grand unification of fundamental forces, then the polarization
signal should be detectable with a next-generation
experiment.  If inflation had something to do with new physics
at much lower energy scales, then the polarization-curl signal could
be much too small to ever be detectable.

Figure \ref{fig:kinney} can be used to make this argument more
precise.  Shown therein are predictions of the amplitude ($r$)
of the gravitational-wave background (measured via the polarization-curl signal)
for five classes of inflationary models.  The four classes with
the largest $r$ arise in models in which
inflation is related to unification of fundamental forces. Generically,
models based on grand unification predict $r$ values in this range. The
fifth class of models (that with the smallest polarization)
usually arises if inflation had something to do with
lower-energy physics.  To rule out all of the grand unification
models shown requires a
polarization sensitivity that will allow detection of a
gravitational-wave amplitude as small as $r\sim0.001$. 
A null result would then indicate
that if inflation occurred, it must have arisen from  new
physics at a lower energy scale.  

To measure $r$ with sensitivity 0.001,
in the presence of foregrounds,
the polarimeter sensitivity required is
$\sim 0.1$~$\mu$K$\sqrt{\rm sec}$, roughly 100 times that of
Planck.  Such a sensitivity level will be
difficult to achieve in the short term.  However, an order-of-magnitude
improvement over Planck sensitivities should be achievable
within the next decade, and this would allow an initial search
for a gravitational-wave background and, in turn, a strong
selection among inflation models.

Foregrounds: Right now we don't know whether our ability to detect gravitational waves will be limited by
detector sensitivity or by foregrounds because the polarization foregrounds
are currently very poorly understood. (\cite{teg99}) Even if MAP and upcoming
ground- and balloon-based experiments do not have sufficient sensitivity 
to detect gravitational waves, they will be able to measure 
the polarization of foreground emission such as the dust grain
emission polarization.  These measurements will be essential in
the design of a future experiment.

Sky Coverage: The current lack of information on polarized foregrounds makes it difficult
to determine the optimum sky coverage for a polarization experiment.
If the subtraction of foregrounds is the main difficulty for a sensitive
polarization experiment, all sky coverage may be essential. High sensitivity is
also needed, however, and better per-pixel sensitivity can be achieved 
by limiting sky coverage.

Site:
A high sensitivity polarization experiment must be done from space for two reasons.
First, the sensitivity required for these observations is not likely
to be possible from within the atmosphere.
Second, uniform coverage over a large region of sky and over a wide range of
frequencies
will be valuable in separating polarized CMB sky structure from
polarized foreground emission. 
Multi-frequency uniform sky coverage was one of the spacecraft advantages 
that placed the COBE results head and shoulders
above previous results. This advantage will also allow MAP and Planck to
succeed where ground and balloon based experiments are struggling. 
(See Figures \ref{fig:spectrum} and
\ref{fig:expplot}.)
The same rules will hold for polarization experiments: 
although good progress will be made from the ground
and from balloons, a spacecraft will eventually be needed.


\section{ Fine Scale CMB Sky Structure }

\begin{figure}
\begin{center}
\BoxedEPSF{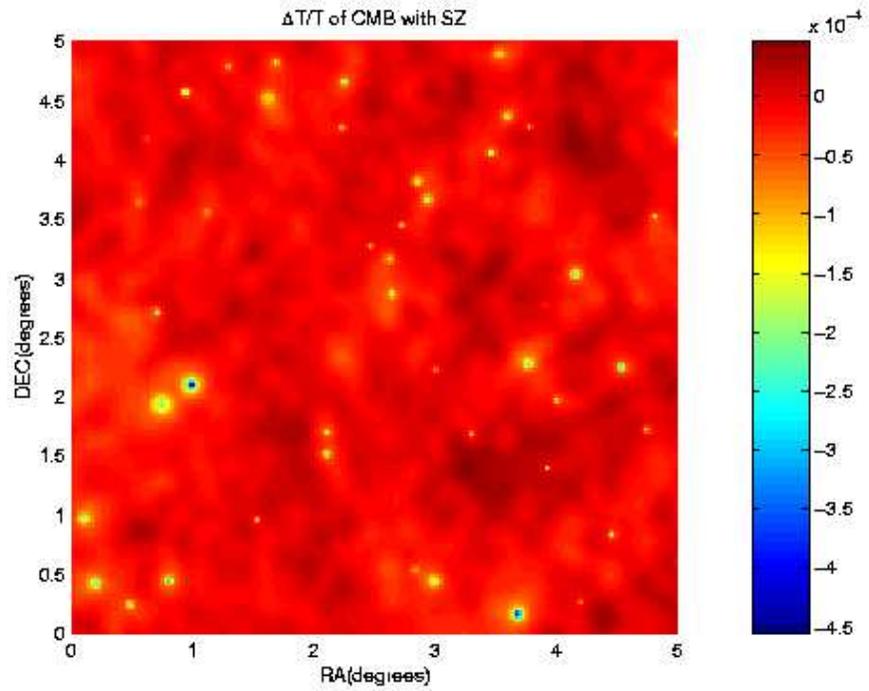}
\end{center}
\caption[Simulated fine angular scale CMB maps]{{\bf Simulated fine angular scale CMB image.\ \ } 
Simulated CMB structure on a five degree square region of sky is shown.
The faint extended structure is primary CMB structure, 
while the
strong localized sources are secondary structure due to
the Sunyaev-Zel'dovich effect in galaxy clusters. At fine angular scales the S-Z signal dominates.
This simulation does not include the S-Z filaments across the sky expected from the collapse of 100Mpc
structures.
\label{fig:cmbszmap} }
\end{figure}

As CMB photons travel from the surface of last scattering
to the observer, secondary sky structure can arise due to the
interaction of the CMB photons with intervening re-ionized matter.
This sky structure is called ``secondary'' in contrast to the
primary structure produced at age 300,000 years.
One interesting and useful source of secondary structure is the Sunyaev-Zel'dovich effect (\cite{sunyaev70}, \cite{sunyaev72}), a distortion of the CMB energy spectrum that
occurs when the CMB photons interact with hot ionized gas.
For example, in a rich cluster of galaxies approximately $10\,$\% of the total mass
is in the form of hot ($\sim 10^8\,$K) plasma.
Compton scattering of CMB photons by electrons in this intra-cluster
(IC) plasma can present an optical depth of $\sim$0.01, resulting in
a distortion of the CMB spectrum 
at the mK level. 
See \cite{SunZel80}; \cite{rephaeli95}; \cite{bir99} for reviews.
Figure \ref{fig:cmbszmap} shows a simulated fine scale observation. On fine scales it is the
thermal S-Z effect that is expected to dominate CMB sky structure.

\ForceWidth{.7\textwidth}

\begin{figure} 
\begin{center}
\BoxedEPSF{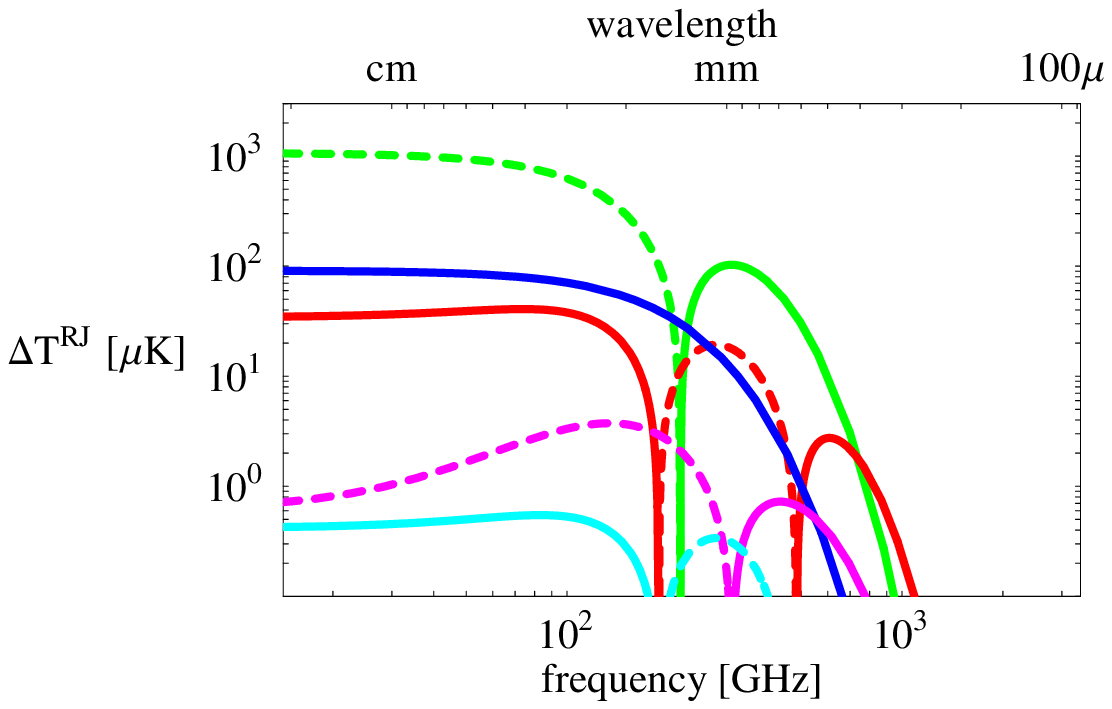}
\end{center}
\caption[The Sunyaev-Zel'dovich Effect]{{\bf The Sunyaev-Zel'dovich Effect} \ \
The Rayleigh-Jeans Temperature (indicating intensity) versus frequency of
different components of the Sunyaev-Zel'dovich effect is plotted for gas in a
fiducial cluster of galaxies. In this simulation, the gas has temperature, $T=10$\,keV; peculiar
velocity, $v=1000km/sec$ towards us; and a Thompson optical depth of
$\tau=0.01$.  Solid curves indicate increments in the CMB temperature, and the
dashed curves decrements. These different components can be differentiated by
their different spectral shape using sufficiently sensitive detectors with good
frequency coverage in the mm and sub-mm bands. From top to bottom the
components are
1) the classical thermal S-Z effect           ($\propto  \tau  T  $) [green];
2) the classical kinetic S-Z effect ($\propto v\tau     $) [blue];
3) relativistic corrections to (1)    ($\propto  \tau  T^2$) [red];
4) thermal corrections to (2)         ($\propto v\tau  T  $) [magenta]; and
5) finite optical depth corrections   ($\propto  \tau^2T^2$) [cyan].
By measuring the amplitude of (1) and (3) one may disentangle $\tau$ and $T$;
the amplitude of (2) additionally gives $v$, however, measurements of this
amplitude will inevitably be uncertain due to the primary CMB anisotropy
behind the cluster which has the same spectral signature. This uncertainty can
be removed by measuring the  amplitude of (4).  A very accurate arcminute
angular resolution map of deviations of the CMB spectra from a blackbody spectrum would provide
information about the internal density, temperature, and velocity structure of
cluster gas at high redshifts.
\label{fig:sz} }
\end{figure}

There are two components of the S-Z effect which result from
distinct velocity components of the scattering electrons.
The {\it thermal} component is due to the thermal (random) motions of the
scattering electrons.
The {\it kinetic} component is due to the bulk velocity of the IC gas with
respect to the rest frame of the CMB.
In Figure \ref{fig:sz}, the spectral distortion produced by each 
of the two components is shown.
As evident from the figure, the two S-Z components have distinct
spectra which can be separated by observations at several millimeter
wavelengths.

\ForceWidth{0.9\textwidth}

When combined with a measurement of
electron temperature, the ratio of the kinetic
and thermal component amplitudes provides a direct measurement of the cluster's
radial peculiar velocity relative to the rest frame of the CMB.
The observed surface brightness difference of both the thermal and kinetic
components is independent of the cluster redshift, as long as the cluster is
resolved.
Clusters are large objects, typically of order $1\,$Mpc, and subtend an
arcminute or more at any redshift.
Therefore, accurate S-Z measurements can be made
throughout the universe, all the way back to
the epoch of formation of the hot IC gas.
A sky survey of S-Z determined velocities would provide us a view of the motions of
test masses (the clusters) throughout the entire Hubble volume, and
show us the evolution of velocity structure over 
much of the history of the universe.
In addition, the evolution of the number density of clusters provides a sensitive test
of gravitational collapse models.

Until recently it has been assumed that X-ray measurements would be needed
to determine the electron temperature of the IC gas.   
But the IC electrons are mildly relativistic, and recent
calculations of the relativistic S-Z effect indicate that with sufficiently
precise CMB measurements it may be possible to determine the electron temperature,
and the cluster velocity, without need for X-ray data (\cite{fabbri81}; \cite{rephaeli95}; \cite{stebbins97}; \cite{challinor98}; 
\cite{itoh98}; \cite{nozawa98}; see Figure \ref{fig:sz}).
This is particularly important for high redshift clusters since the received X-ray flux 
falls as $\sim(1+z)^{-4}$
while the S-Z brightness is redshift independent.  Of course, for nearby
clusters, for which X-ray data is available, both techniques should be used and the
results compared.

In the last few years, high signal-to-noise detections and images have
been made of the thermal S-Z effect toward several distant clusters ($z>0.15$).
Most of these observations have been made
at centimeter wavelengths; observations at $2\,$mm, however,
 have also been successful (\cite{bir99}).
All of these observations have been done using ground based telescopes.

The thermal S-Z distortion is a measure of the thermal history of
the universe. There are several potential sources of heating for the
intergalactic medium: gravitational collapse, energy input
from galaxy superwinds (\cite{pen99}), and energy input from
quasars and AGNs (\cite{nata99}). Since the S-Z effect depends upon $n_e$ rather than $n_e^2$,
we can use it to trace the thermal history not only of dense clusters
but also of filaments. If the mean electron temperature is
1 keV, then a 1 Mpc wide filament produced by the collapse of
a 10 Mpc wave should produce a distortion of 10 $\mu$K.

These filaments, one arcminute wide and about a degree long, 
can be detected by searching for the S-Z
thermal effect distortion they produce on the sky.  The filaments 
will be difficult to detect through
x-ray emission because the electron density in the filaments is too low;
however, the S-Z thermal effect signal they produce should be detectable.

The thermal S-Z effect is expected to be the strongest source of sky structure
at arcminute angular scales but other sources of structure can also 
provide information from the dark ages.  
Once arcminute resolution multi-frequency CMB maps are available, the regions 
affected by the thermal S-Z effect can be identified. S-Z free regions can then be
studied to detect other sky structure present.  The physics responsible
for fine scale structure include gravitational lensing,
bulk motions of plasma (the Ostriker-Vishniac effect),
evolution of the gravitational potentials during the passage of CMB photons (the Integrated Sachs 
Wolfe effect and the Rees-Sciama effect), and details of the ionization history of the universe. 
See \cite{hu97a}, \cite{refregier98}, and references therein for details.

\smallskip
\smallskip
\centerline{\bf Specifications of Fine Scale Observations}
Angular resolution: One arcminute resolution will be required.
At high redshifts the cores of galaxy clusters subtend about one arcminute.
In addition the S-Z filaments produced by gravitational collapse 
should be about an arcminute across, and as long as a degree.
At the millimeter wavelengths needed for CMB observations, arcminute
beamwidths translate to a telescope aperture (or longest baseline) of 
about 10 meters.

Sensitivity:  The brightest clusters can produce S-Z   amplitudes
$\sim$1mK. The kinetic effect is expected to be smaller, $\sim$200 $\mu$K.  Filaments are
expected to produce 10 $\mu$K thermal S-Z signals.
Currently CMB instruments are measuring the $\sim$100 $\mu$K
degree scale sky structure with 5 $\mu$K precision, so
current technology is sufficient for S-Z cluster work. Better sensitivity will
be needed to study filaments.

Frequency Range: To separate S-Z thermal, kinetic, and relativistic-electron effects,
maps covering the range from 30-400 GHz will be needed.
Ideally all these maps should be made using the same instrument.
Additional information at lower (5-10 GHz) frequencies and higher 
(1000-3000 GHz) 
frequencies will also be needed as an aid in removing
emission from unrelated foreground and background sources.

Telescope technology: Both filled aperture and interferometric telescopes can be used for
observations at this angular scale and frequency.  The advantages of interferometers
include:  rejection of telescope emission, insensitivity to amplifier gain fluctuations,
and fault tolerance.
The advantages of filled aperture telescopes include: wide detection bandwidth,
wide observing frequency range, lower complexity, and lower cost.

\begin{figure}
\begin{center}
\BoxedEPSF{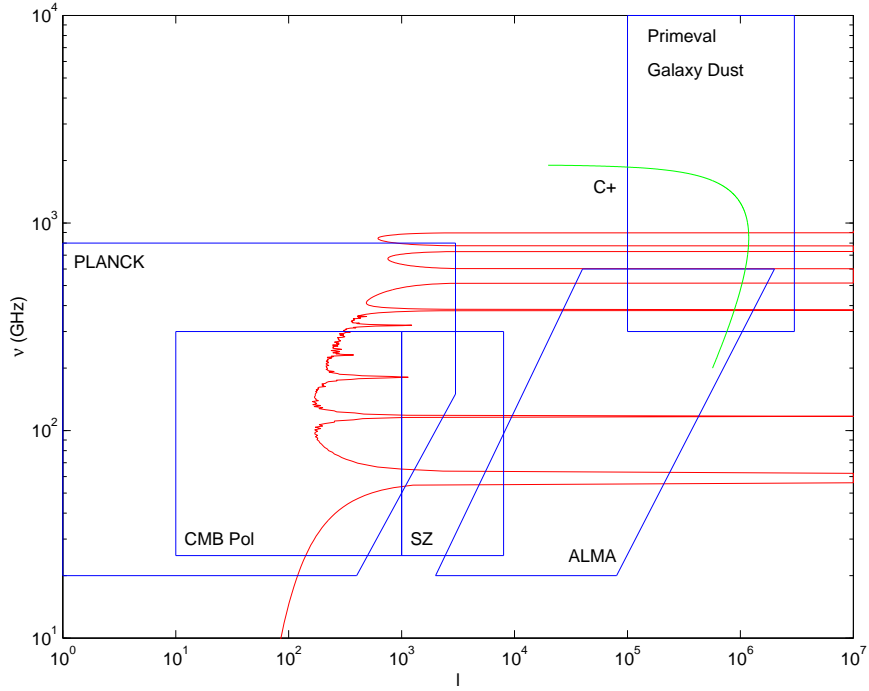}
\end{center}
\caption[Comparison of Experiments]{{\bf Comparison of Experiments\ \ } 
Shown in angular scale $\ell$ versus observing frequency $\nu$ space are science targets:
CMB polarization, Sunyaev-Zel'dovich effect, dust emission by primeval galaxies, and
the fine structure line of ionized carbon from a high redshift galaxy. Also shown are the
domains of two instruments: the Planck spacecraft and the Atacama Large Millimeter Array (ALMA).
(\cite{alma99}) The red jagged curve is a comparison of
atmospheric noise to noise in a spacecraft
environment. On this curve the
South Pole winter atmospheric emission fluctuations equals the photon 
fluctuations in a spacecraft environment (assumed due to the brightness of the sky along 
with photons emitted from a 1\% emissivity 70 K surface).
Observations above and to the left of this curve derive a noise benefit from being in space.
Cluster S-Z observations fall to the right of the noise boundary, 
and are appropriate for ground based
observation, while CMB polarization observations should be done from space.
\label{fig:expplot} }
\end{figure}

Site: As shown in Figure \ref{fig:expplot}, observations of the
thermal S-Z effect in clusters should be 
attempted from the ground before resorting to spacecraft
observations. 
Currently ground based efforts are making rapid progress on this topic.  However, attempts
to measure electron temperatures in clusters via the relativistic S-Z effect 
and attempts to map peculiar velocities over large regions of sky
may require spacecraft observations.
S-Z filament observations, which require multi-frequency 
maps covering degrees, along
with microkelvin sensitivity, may also benefit from observations done in space.

\ForceWidth{.7\textwidth}

\newpage
\section{ Summary }

After the Planck mission, we will need to make precise 
multi-frequency, all-sky maps of CMB polarization.
CMB polarization measurements, because they may allow for the detection
of a gravitational wave background, probe much further back in time
than any  other astronomical observation.  In so doing, CMB polarization
measurements allow the study of particle physics at energies far higher than are
available with any earth-bound accelerator experiment.  
The processes that may have produced 
this background of gravitational waves probably involve the unification of the
the strong and electroweak forces and may also involve quantum gravity.
Because of the ability to constrain physics at
enormous energy scales, CMB polarization observations have the potential to revolutionize
our understanding of basic physics.  Polarization information 
is also expected to help determine the class of primordial perturbations
(density; pressure; gravitational wave), and will allow a
strong selection among inflation models.

After Planck it will also be important to make arcminute angular scale 
observations of the CMB because these observations can probe the dark ages of
cosmology.  Using fine scale CMB observations we can see 
the largest structures of the 
universe, galaxy clusters and galaxy sheets, as they were first forming.

This committee has sought comment from over one hundred cosmologists, as well as 
from members of the particle physics and general relativity communities.
The discussion has been wide-ranging, but the same comment was heard 
again and again.  
Because of the potential for
discovery of dramatic new physics (gravitational waves from inflation),
along with model-testing power, CMB polarization measurements
should have the highest priority among CMB observations following
Planck.

It would be truly remarkable if, early in the twenty-first century,
we could say that we understand what was happening in the universe
just $10^{-38}$ seconds after it began.

\newpage
\section{Acknowledgements}

Chris Cantalupo produced Figures \ref{fig:cmbszmap} and \ref{fig:expplot}. Greg Griffin,
Michael Vincent, Michael O'Kelly, Kurt Miller, and Gabrielle Walker provided editorial comments. Greg Wright carried out atmospheric emission calculations.

%
%
%
%

{
\protect\itemsep=0pt
\protect\parsep=0.1\baselineskip
\bibliographystyle{aas}

\begin{thebibliography}{}

\bibitem[Abbot and Wise 1984]{AbbWis84} Abbott, L.~F. and Wise,
     M., {\em Nucl. Phys.} {\bf B244}, 541 (1984)

\bibitem[Bennett et al. 1996]{bennett96}
Bennett, C.~L. et al. \apj, {\bf 464}, L1. (1996)

\bibitem[Birkinshaw 1999]{bir99} Birkinshaw, M.  {\it Phys.\ Rep.}, {\bf310}, 97. (1999)

\bibitem[Bond and Efstathiou 1984]{BonEfs84} Bond, J.~R. and Efstathiou, G. \apjl, {\bf285}, L45. (1984)

\bibitem[Caldwell, Kamionkowski, and Wadley 1998]{caldwell98}
Caldwell, R.~R., Kamionkowsi, M., and Wadley, L.  \prd, {\bf 59} (1996), astro--ph 9807319

\bibitem[Challinor and Lasenby 1998]{challinor98}
Challinor, A. and Lasenby, A. \apj, {\bf 499}, 1. (1998)

\bibitem[Fabbri 1981]{fabbri81} Fabbri, R. \apss, {\bf 77}, 529. (1981)

\bibitem[Fixsen et al. 1994]{Fixetal94} Fixsen, D.~J., Cheng, E.~W., Cottingham,
 D.~A., Eplee, R.~E., Isaacman, R.~B. 
et al. \apj, {\bf402}, 32. (1994)

\bibitem[Halpern and Scott 1999]{halp99}
Halpern, M. and Scott, D. to appear in {\it Microwave Foregrounds}, 
eds. A. de Oliveira-Costa \& M. Tegmark (ASP, San Francisco), (1999), 
astro--ph/9904188

\bibitem[Hu and White 1997a]{hu97a}Hu, W. and White, M. \apj, {\bf497}, 568. (1997)

\bibitem[Hu and White 1997b]{HuWhi97b} Hu, W. and White, M. {\it New Astron.}, {\bf2}, 323. (1997)

\bibitem[Itoh, Kohyama, and Nozawa 1998]{itoh98}
Itoh, N., Kohyama, Y., and Nozawa, S. \apj, {\bf 502}, 7. (1998)

\bibitem[Kaiser 1983]{kaiser83}Kaiser, N. {\it MNRAS}, {\bf202}, 1169. (1983)

\bibitem[Kamionkowski, Spergel, and Sugiyama 1994]{kamionkowski94}
Kamionkowski, M., Spergel, D.~N., and Sugiyama, N. \apj, {\bf426}, (1994), astro--ph/9401003

\bibitem[Kamionkowski, Kosowsky, and Stebbins 1997]{KamKosSte97a}
Kamionkowski, M., Kosowsky, A., Stebbins, A., Phys. Rev. Lett. {\bf 
78}, 2058 (1997)

\bibitem[Kamionkowski 1998]{kamionsci98}
Kamionkowski, M. {\it Science} {\bf280}, 1397, (1998)

\bibitem[Kamionkowski and Kosowsky 1998]{kamionkowski98}
Kamionkowski, M. and Kosowsky, A. \prd, {\bf 57}, 685. (1998)

\bibitem[Kamionkowski and Kosowsky 1999]{KamKos99} Kamionkowski,
M. and Kosowsky, A.  To appear in {\em
Annu. Rev. Nucl. Part. Sci.} (1999), astro-ph/9904108

\bibitem[Keating et al. 1998]{keating98}Keating, B., Timbie, P., Polnarev, A., and
Steinberger, J. \apj, {\bf495}, 580. (1998)

\bibitem[Kinney 1998]{kinney98}
Kinney, W.~H. \prd {\bf 58}, 123506, (1998), astro--ph 9806259.


\bibitem[Kogut et al. 1995]{kogut95}
Kogut, A., Banday, A.~J., Bennett, C.~L., Gorski, K.~M., Hinshaw, G., and Reach, W.~T. (1995)
astro--ph/9509151

\bibitem[The Millimeter Array Web Site 1999]{alma99}
The Millimeter Array Web Site. http://www.mma.nrao.edu/ 01 June 1999

\bibitem[Natarajan and Sigurdsson 1999]{nata99}
Natarajan, P. and Sigurdsson, S. {\it MNRAS} 302,288. (1999)

\bibitem[Nozawa, Itoh, and Kohyama 1998]{nozawa98}
Nozawa, S., Itoh, N., and Kohyama, Y. \apj, {\bf 508}, 17. (1998)

\bibitem[Pen 1999]{pen99}
Pen, U., to appear in {\it Astr J Lett}, (1999), astro--ph/9811045

\bibitem[Polnarev 1985]{polnarev85} Polnarev, A. {\it Sov Astron}, {\bf29}, 607. (1985)

\bibitem[Rees 1968]{rees68}Rees, M. {\it Astr J Lett}, {\bf153}, L1. (1968)

\bibitem[Refregier, Spergel, and Herbig 1998]{refregier98}
Refregier, A., Spergel, D.~N., and Herbig, T., (1998), astro--ph/980634 

\bibitem[Rephaeli 1995]{rephaeli95}
Rephaeli, Y. \apj, {\bf445}, 33. (1995)

\bibitem[Sachs and Wolfe 1967]{SacWol67} Sachs, R.~K., Wolfe,
A.~M., \apj, {\bf 147}, 73 (1967)

\bibitem[Seljak and Zaldarriaga 1997]{Zal97} Seljak, U. and
Zadarriaga, M.  Phys. Rev. Lett. {\bf 78}, 2054 (1997)

\bibitem[Smoot et al. 1992]{Smoetal92} Smoot, G.~F., Bennett, C.~L., Kogut, A., Wright, E.~L., Aymon, J., et al. \apjl, {\bf396}, L1. (1992)

\bibitem[Stebbins 1997]{stebbins97}
Stebbins, A. (1997) in {\it The Cosmic Microwave Background} ed.s C.H.Lineweaver,
J.G. Bartlett, A. Blanchard, M. Signore, \& J. Silk (Dordrecht: Kluwer)

\bibitem[Sunyaev and Zel'dovich 1970]{sunyaev70}
Sunyaev, R.~A. and Zel'dovich, Y.~B.
{\it Comments Astrophys. Space Phys.}, {\bf2}, 66. (1970)

\bibitem[Sunyaev and Zel'dovich 1972]{sunyaev72}
Sunyaev, R.~A. and Zel'dovich, Y.~B.
{\it Comments Astrophys. Space Phys.}, {\bf4}, 173. (1972)

\bibitem[Sunyaev and Zel'dovich 1980]{SunZel80} Sunyaev, R.~A. and Zel'dovich, Y.~B. {\it ARAA}, 
{\bf 18}, 537. (1980)

\bibitem[Tegmark et al. 1999]{teg99}
Tegmark, M., et al., (1999), astro--ph/9905257

\bibitem[White, Scott, and Silk 1994]{WhiScoSil94}
White, M., Scott, D., and Silk, J. {\it ARAA}, {\bf32}, 319. (1994)





\end{thebibliography}
}
{}

\clearpage

\end{document}